\documentclass{article}

\usepackage{arxiv}

\usepackage[utf8]{inputenc} 
\usepackage[T1]{fontenc}    
\usepackage{hyperref}       
\usepackage{url}            
\usepackage{booktabs}       
\usepackage{amsfonts}       
\usepackage{nicefrac}       
\usepackage{microtype}      
\usepackage{lipsum}
\usepackage{graphicx}
\usepackage[section]{placeins}

\title{Comparison Study for Multi-Vendor vs Single-Vendor for Enterprise Computer Networks}

\author{
  Edmond Shami\\
  Telecommunications Design Engineer\\
  Dar Al-Handasah\\
  Amman, Jordan \\
  \texttt{edmond.shami@gmail.com} \\
   \And
 Abdelmalek Saleh \\
  Telecommunications Design Engineer\\
  Dar Al-Handasah\\
  Amman, Jordan\\
  \texttt{abdelmalek.hirzallah@gmail.com} \\
}

\begin{document}
\maketitle

\begin{abstract}
One of the topics that concerns the way computer networks are designed, is the single-vendor and multi-vendor solutions. Where the performance and operation of your network depends on which model you choose for your enterprise, and the future risks aligned with such models. This study outlines the strengths and average price ranges of multiple vendors in the past 2 years (2018 and 2019), practical cases in which each model works, a case study done by Gartner, and finally, recommendations that can help push the design practices when it comes to network design. 

\end{abstract}


\section{Introduction}
Computer networks are one of the most, if not the most, critical systems among telecommunication systems for enterprise buildings. The system is considered the backbone for other systems and the one that interconnects them together, not forgetting the critical role it plays in business operations as the services it offers go beyond day-to-day communications between users. That’s why, it needs to get the attention it deserves in terms of staying up-to-date with network trends and best practices.

\section{Vendors: Strengths and Price Ranges}
Networks are made up of multiple components (core switching, edge switching, firewalls, wireless access points, etc...) and there is no single company that offers the best of each component in the market. Hence, enterprises, depending on their size and environments in which they operate, tend to obtain the best components (in terms of performance, price and maintenance services) from multiple vendors or obtain the whole solution from a single vendor with the best that single vendor can offer. 

Figures 1, 2 and 3 illustrate the strengths of each vendor in the main categories in networking systems, along with the price range compared to other vendors.

\begin{figure}[htp]
    \centering
    \includegraphics[width=17cm]{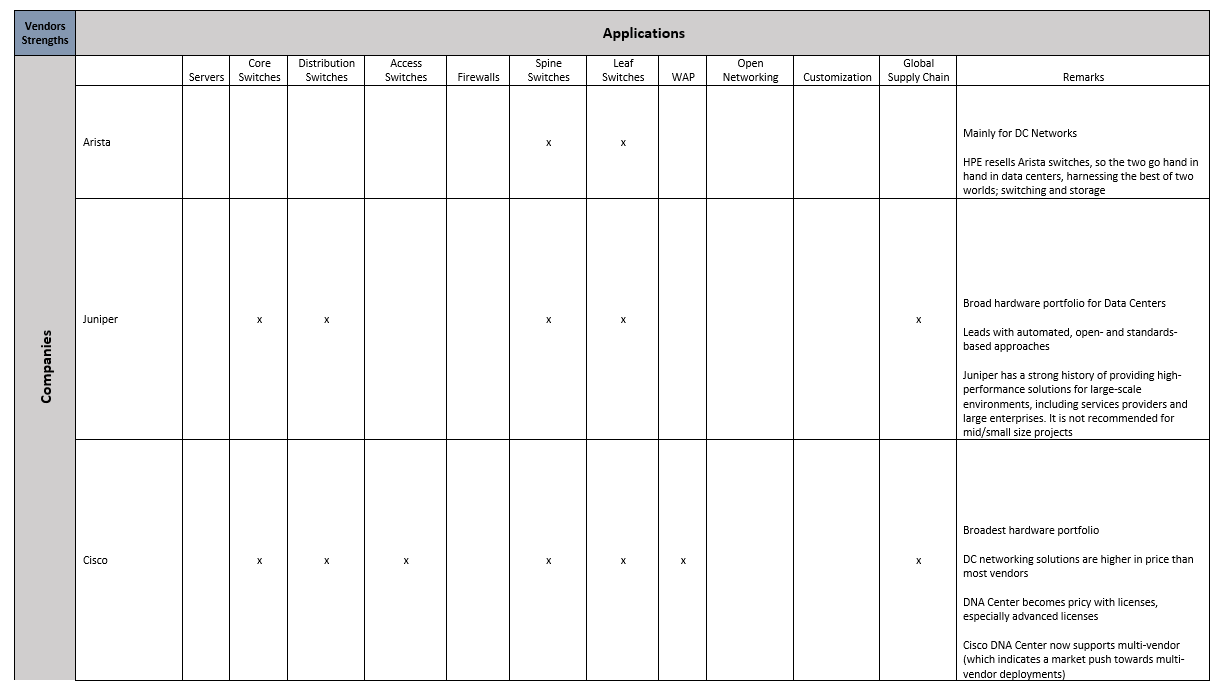}
    \caption{Vendors Strengths 1/2 [1] [3]}
    \label{fig: strength}
\end{figure}

\begin{figure}[htp]
    \centering
    \includegraphics[width=17cm]{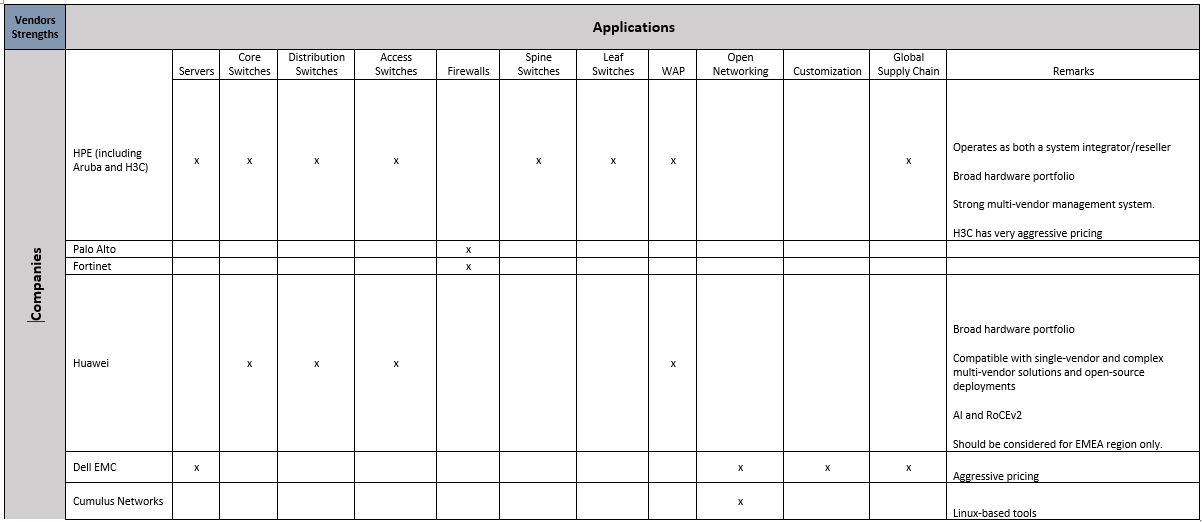}
    \caption{Vendors Strengths 2/2 [1] [3]}
    \label{fig: strength 2}
\end{figure}

\begin{figure}[htp]
    \centering
    \includegraphics[width=15cm]{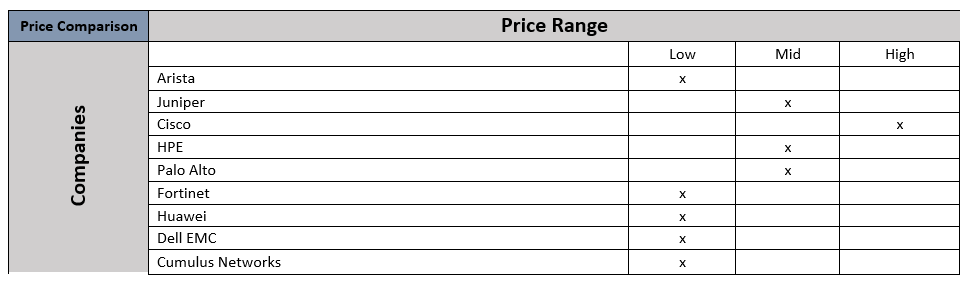}
    \caption{Relative Price Range of Each Vendor}
    \label{fig: price}
\end{figure}

\section{Single-Vendor VS Multi-Vendor}
Each client has their own network requirements, budget constraints, and risks which they can afford, all of those depend on the client’s holdings size and the area in which they operate. Hence, when it comes to the selection between single or multi-vendor design, the client and consultant should find the best solution which suits the client needs according to Figure 4 below. 

\begin{figure}[htp]
    \centering
    \includegraphics[width=18cm]{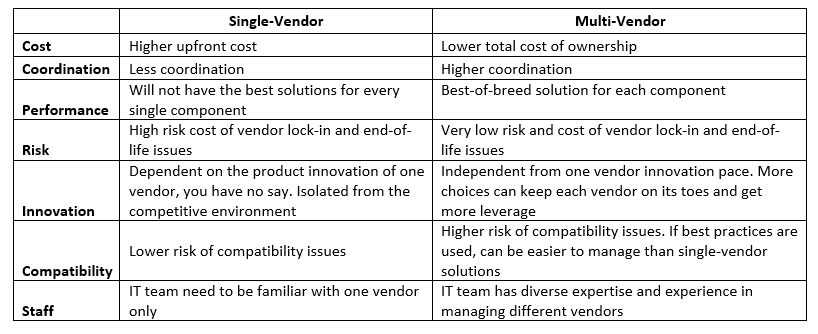}
    \caption{Final Comparison}
    \label{fig: comparison}
\end{figure}

\section{A Case Study by Gartner}
The following cost case study (Figure 5) was done by Gartner of an organization replacing a network with 100 to 200 access switches and the associated aggregation and core switches.

\begin{figure}[htp]
    \centering
    \includegraphics[width=15cm]{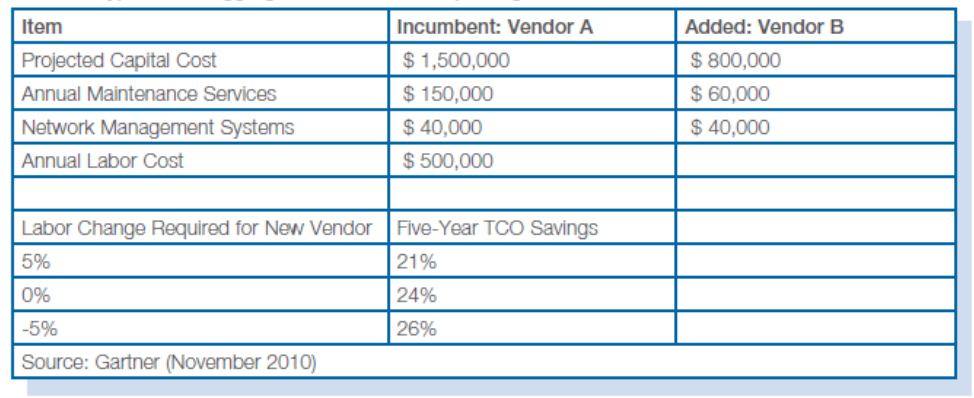}
    \caption{Typical and Aggregated Results for Replacing a Network with 100 to 200 Access Switches [6]. **Maintenance would be the next business day for edge products with an on-site, four-hour response for the core network}
    \label{fig: case study}
\end{figure}

As can be observed, a two-vendor solution had a drastic cost reduction, and surprisingly, not only in capital costs, but also in maintenance costs.

Gartner found out that single-vendor solutions enable the vendor to demand a premium on maintenance costs, while there are vendors who demand less maintenance costs and in some cases, lifetime warranties, such as HPE. 

In one of Gartner’s surveys, Organizations achieved capital cost savings of 30\% to 50\% and 40\% to 95\% less than Cisco's SmartNet services in terms of warranties and maintenance. From the same survey, it was found that no additional staff is needed. And interoperability issues are easily managed. In fact, multi-vendor solutions become easier to manage in the long run.

\section{Conclusions and Recommendations}

Moving forward, and based on the data presented above, the following recommendations are proposed to enhance the design guidelines:

\begin{itemize}
\item If you don’t deploy a multi-vendor solution in the early design stages, you need to at least make the network supportive of future multi-vendor solutions. Which can also be included in the specifications, since most of customers surveyed indicated that multivendor applications were a highly important factor in their vendor selection criteria.
\item There are best practices that can be used in a multi-vendor environment that further reduces any compatibility issues. Which can also be added to specifications, if the option of the multi-vendor is included. These include; having well defined boundaries for the active equipment (edge switching, core switching, processing equipment, server environments, etc.), reducing interface points to a minimum and not randomly mixing products,and ensure that you are following international standard and not specifying any proprietary based systems/protocols.

\item A distinction shall be made between two main networks: Data Center and Campus Network. Each network has its own characteristics. Data Centers, especially the big ones, tend to favor multi-vendor solutions. On the other hand, campus networks, small to mid-size ones can handle a single-vendor solution, but it can also depend on the risks and cost the client is willing to take and the environment in which they operate and whether or not they have any good established relationships with one of the vendors.

\item The Spine-Leaf topology is now the most common physical network design in Data Centers, proposed by most vendors, it has replaced the  three-tier design (access, distribution, and core). It is recommended to make the specifications compliant with such a topology in terms of hardware, associated protocols, and design criteria.

\item Sometimes the client requirements do not require the best of breed, but the best price for a certain technical requirement. It is not a necessity to always specify the top-tier equipment everywhere, as an example the CCTV Campus network does not have the same requirements as the ICT network, as such lower your specifications to match your customer needs keeping in mind any future requirements.

\end{itemize}

It is worth mentioning that the networking giants are following an approach of either acquiring new and upcoming companies that excel in making a certain product (Cisco acquisition of Meraki, Firepower, etc.), or by collaborating with other already big names in the market (HPE collaborating with Arista, Aruba, etc.) which is additional proof of the strengths of multi-vendor solutions as opposed to a single-vendor solutions.

\bibliographystyle{unsrt}  


\end{document}